# Suicide Classificaction for News Media Using Convolutional Neural Network

**Hugo J. Bello[1,*], Nora Palomar-Ciria[2], Enrique Baca-García[3-10], and Celia Lozano[11]**

[1]*Department of Applied Mathematics, Universidad de Valladolid, Soria, España.*
[2]*Servicio de Psiquiatría, Complejo Asistencial de Soria, Soria, España.*
[3]*Department of Psychiatry, Hospital Universitario Fundación Jiménez Díaz, Madrid, Spain.*
*INSERM Unit 1061, Montpellier, France; CHU Nimes and University of Montpellier, France.*
[4]*Department of Psychiatry, Universidad Autónoma, Madrid, Spain.*
[5]*Department of Psychiatry, Hospital Universitario Rey Juan Carlos, Móstoles, Spain.*
[6]*Department of Psychiatry, Hospital General de Villalba, Madrid, Spain.*
[7]*Department of Psychiatry, Hospital Universitario Infanta Elena, Valdemoro, Spain*
[8]*CIBERSAM (Centro de Investigación en Salud Mental), Carlos III Institute of Health, Madrid, Spain.*
[9]*Universidad Católica del Maule, Talca, Chile.*
[10]*Department of Psychiatry, Nimes University Hospital, Nimes, France.*
[11]*Department of Big Data and Business Intelligence, Sermes CRO, Madrid, Spain.*

***Corresponding author. Email:** hugojose.bello@uva.es **(Hugo J. Bello)**

**ABSTRACT**

**Currently, the process of evaluating suicides is highly subjective, which limits the efficacy and accuracy of prevention efforts. Artificial intelligence (AI) has emerged as a means of investigating large datasets to identify patterns within 'big data' that can determine the factors on suicide outcomes. Here, we use AI tools to extract the topic from (press and social) media text. However, news media articles lack of suicide tags. Using tweets with hashtags related to sucide, we train a neuronal model which identifies if a given text has a suicidade-related contagion. Our results suggest a high level of the impact of mediatic into suicide cases, and a intrinsic thematic relationship of suicide news. These results pave the way to build more interpretable suicide data, which may help to better track, understand its origin, and improve prevention strategies.**

## 1. INTRODUCTION

Suicide is a major public health issue around the world, being the second cause of death in young people (especially under 24 years) (*Suicide Data Report - 2012*, n.d.; *WHO | Preventing Suicide*, n.d.-a). Spain ranked 130th of the 170 countries listed according to WHO's 2016 suicide report (Suicide, 2019; Wolfram Research, Inc., n.d.). A particular suicidade case in Spain is "Suicides for eviction", i.e. linked to issues relating to housing and mortgage debt, where 34% of the suicides result from them (El 34 Por Ciento de Los Suicidios Que Se Producen En España Son Por Los Desahucios – Alerta Digital, n.d.). Sometimes the link between eviction and suicide is unknown, and as in most suicides, is due to complex and difficult personal situations. Many factors have been studied to have an impact on suicidal ideation, and among them contagion effect (poverty, in family, in celebrities and in press) has been a topic of interest for years (Carmichael & Whitley, 2018). On this issue, media covering of suicide has been studied to be influential, not only in a negative way but also (and more recently studied) it can have a protective effect(Chan et al., 2003; Niederkrotenthaler et al., 2010b). Harmful influence on the reporting of suicide in media is known as Werther effect and it worsens with the content and the way suicide death is depicted(Phillips, 1974; Sisask & Värnik, 2012; Stack, 2002, 2003). On the contrary, Papageno effect is a protective effect on suicidal behavior determined by overcoming facts on suicide and empathetic responses to people suffering suicidal ideation (Arendt et al., 2019; Carmichael & Whitley, 2018; Niederkrotenthaler et al., 2010a, 2010b; Rosen et al., 2019).

News media are key to understanding how the population reacts to a certain topic (Dijk, 1995). The role of suicide-related media have led to the creation of guidelines for media reporting. Indeed, the topic of news can influence the public perceptions. Because of that not only news media convey what happens, but also what could be done about it. Recently, Marchant et al. (Marchant et al., 2017) evidenced that the Internet and social media can influence suicide-related behavior. Indeed, analyzing Tweets can identify individual human beings' opinions.

The main aim of this research is to get a better knowledge of suicides-related contagion publishing in Spanish Media using Artificial Intelligence (AI). At the population level, algorithms can identify the main topic of a text, predict the risk and develop prevention guidance. Here, we use Machine Learning techniques, i.e. Natural Language Processing (NLP) (Chapman et al., 2011), to analyze high amounts of tweets sent by individual persons to create a text classification model for suicidal connotation of any text. To date, the efforts on NLP involve topic identification, natural language understanding, and natural language generation (Neural Networks for Natural Language Processing de S., Sumathi, M., Janani - Libros En Google Play, n.d.). To obtain the main subjects of written media content, we use the trained neural networks. Then, we search for topics related to suicidade and see how they are reported and how they are related. Our findings show two aspects: (i) the global impact of mediatic suicide cases, (ii) the intrinsic thematic relationship of suicide news.

## 2. MATERIAL AND METHODS

### 2.1 Data Extraction and Topic Classification

We use a Big Data technique called web scraping (also called web crawling or mining) to extract such an immense amount of texts. In particular, this technique gets a group of servers querying massive quantities of data from public web pages on the internet. We deploy a network of cloud servers and a local server to break up the load of the work: (i) A Mongodb database server: a no-sql database widely used in Big Data projects in order to store large amounts of data. (ii) Two Cloud servers (scrapers): these network nodes contain our software coded with the Golang programming language. This software searches the web archive of the newspapers, then the nodes save the data in the Mongodb database. (iii) A local server: A server with scripts coded in NodeJS and Python searches our Mongodb database and trains the neural networks to classify the subjects in the news. Then, this server analyzes the full database and by using neural network models calculates the probability that each new has of covering the subject of suicide.

We extract 784 259 news from Spanish newspapers (from lavanguardia, elpais, elmundo, abc, 20minutos, publico and diario.es) from January 2005 to March 2020. In the media, every piece of news is classificated by tags. This tag or keyword helps describe the text and allows it to be found again by browsing or classificating (such as *justice, economy, politics, international, technology, health*,...) (see the most frequent tags in Fig.1a). News media articles, however, lack of suicide tags, which hamper the correponding suicidade analysis. To overcome such issues, we use Twitter texts. In Twitter, people use the hashtag symbol (#) before a relevant keyword in their tweet to categorize those tweets and help them show more easily in Twitter search. We extract over the same period of time 143 160 tweets in spanish with hashtags related to sucide, i.e. "#suicidio"(suicide), "#suicida", "#suicidó" (died by suicide), "#suicidaron" (died by suicide), "#suicidando" (suiciding), "#autolesión" (self harm), "#suicidioinfantil" (infantile suicide), "#suicidioadolescente" (teen suicide), "#suicidiojuvenil (young suicide). Firstly, we select all these tweets to train a topic classification model, i.e. a multilabel subject classification NN1 model. Due to this subgroup of texts has been already suicidade tagged, it allows the neural network to learn from them. NN1 consists of a convolutional neural network composed of an embedding layer, two convolutional (1 dimensional) layers, one pooling and one output dense layer. This neural network was trained with binary cross entropy loss of 0.0153 (multilabel accuracy 15%). Given a text input (news content for instance) this neural network outputs a set of tags (subjects) with their probabilities. From such classification, for each new we obtain a probability (number between 1 and 0) that represents how much this new covers the subject of suicide (a sketch of this process is depicted in Figure 1b).

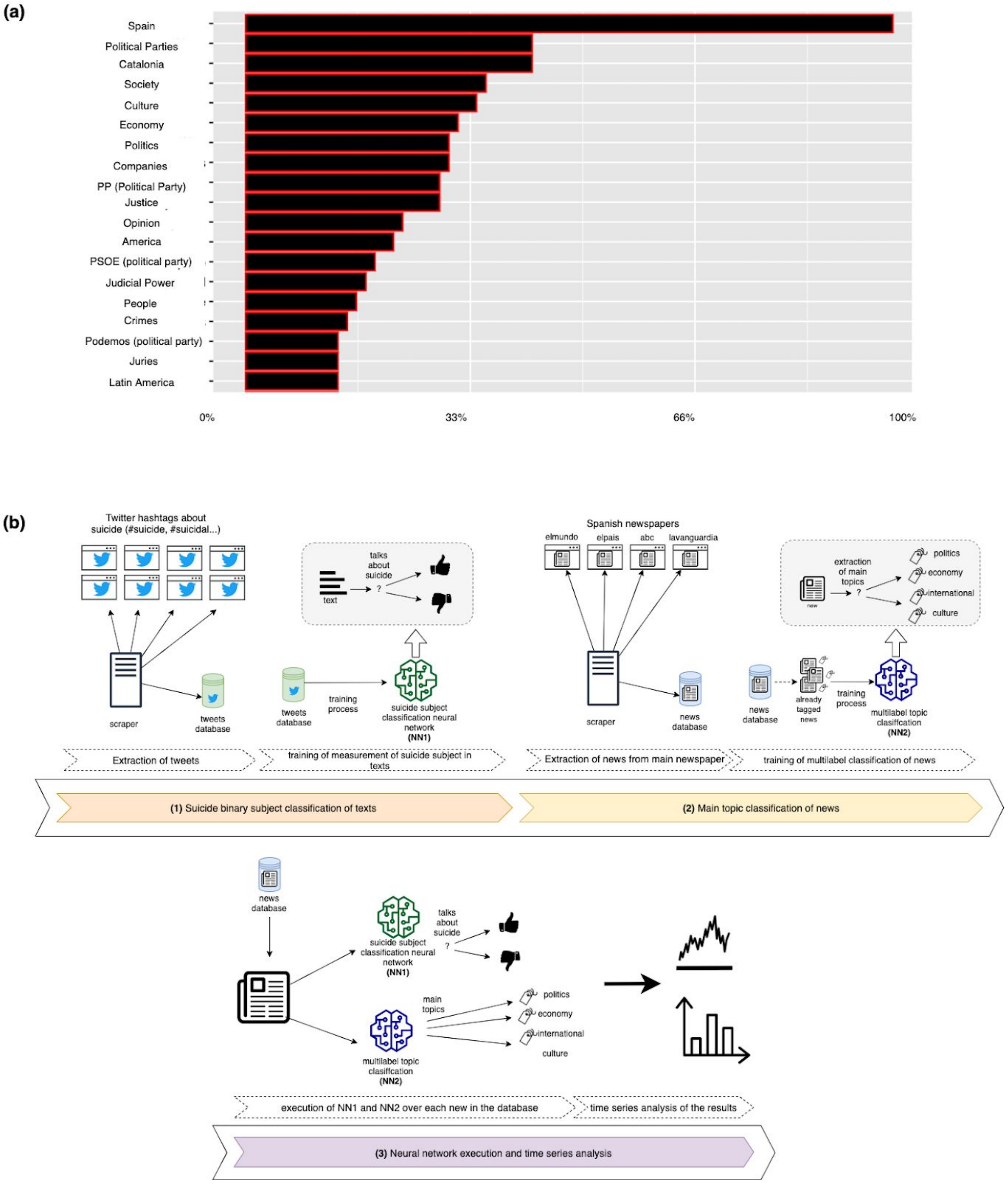

**Figure** 1. (a) Top 20 Topic Classification of news article. (b) Workflow of the topic classification process. On the left, the data extraction and training of a neural network (NN1) capable of determining if a text talks about the subject of suicide. On the right, the data extraction and training of a neural network (NN2) capable of obtaining the main subjects of a new (such as politics, international, economy ...). On the bottom, the summary of the applied process: firstly, we apply each neural network over all the news of the database and secondly, analyze the results using time series.

## 3. RESULTS

### 3.1 The impact of mediatic suicide cases

All results presented in the following were carried out for the news media topic classification (obtained with the model NN1 described in the previous methods). To understand the press coverage of suicide in more detail, we analyze the suicide subject probability over time, i.e. the temporal changes (*Modelling Stationary Time Series: The ARMA Approach | SpringerLink*, n.d.). Considering the whole daily news, we measure the temporal daily average of suicide subject probability from 2005 to 2020 (see Fig. 2a). The larger the probability, the greater its chances to talk about suicide news on a certain day. As we can see, this average probability does not follow a clear trend, but rather sporadic highs and lows over time. In particular, there is an increment of the suicidades (i) since the start of the economic crisis 2010 and 2014, (ii) around 2015-2017, there is a surge deployed in both the number of suicide bombings , (iii) in 2020, due to coronovirus pandemic. To characterize the time variance, we have extracted: (Fig. 2b) the weekly seasonality associated with mediatic suicidades : more suicide subject probability on the day of the weekend (Saturday and Sundays) and Fridays; and (Fig. 2c.) shows a monthly seasonal component modeled using Fourier series, where winter months and August get higher probability.

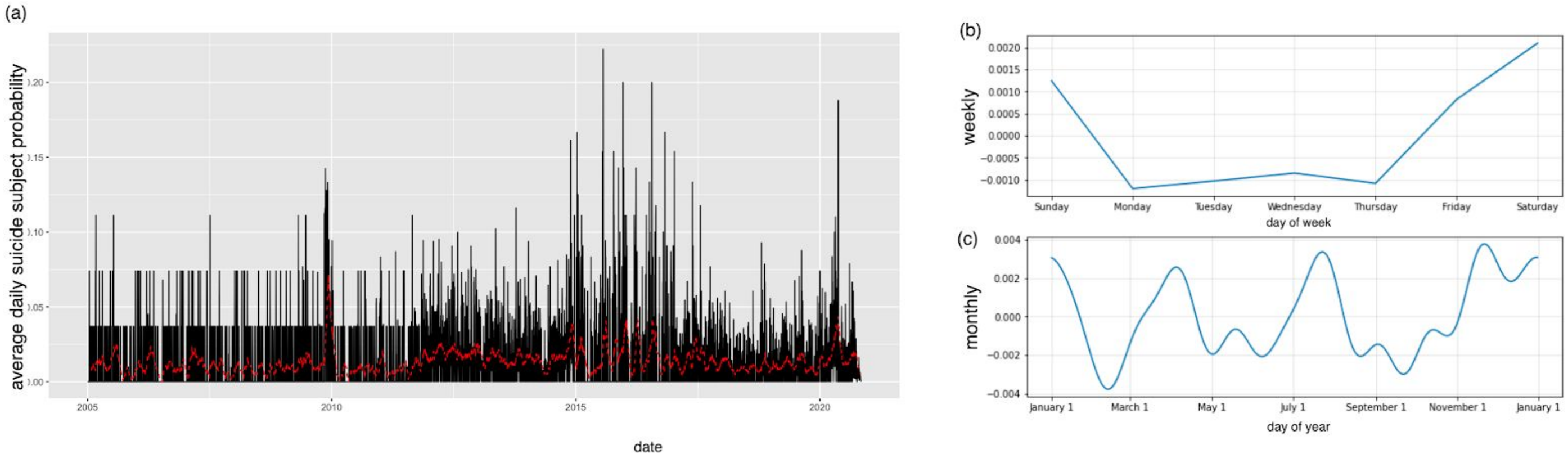

**Figure 2**. (a) average value of suicide subject probability on each day (black lines) and moving average smoothing(red line). (b) the week seasonality and (c ) the month seasonality of the series.

In order to get more insights, we now focus on the daily suicide subject probability by decomposing the time series $Y_t$ with three main model components (Harvey & Shephard, 1993):

$Y_t=T_t+S_t+H_t+e_t$ (2)

where $T_t$ is the trend, $S_t$ is the seasonal component, $H_t$ is the holiday component and $e_t$ is the error. To fit the data, we use the Prophet model (a facebook open software algorithm) (Shen et al., 2020) , in which $T_t$ can be either a rating growth model, or a piecewise linear model, $S_t$ is built using Fourier series and $H_t$ is set using the known holidays. Figure 3 shows the Prophet fit (green lines) of the suicide subject daily data (black symbol). Although it seems that the data fits well in the plot, certain days are out-of-the-range, which represents anomalies in the data trend. Conversely, every time value outside a 99% uncertainty interval of the fitting-model is considered an anomaly (represented as red symbols in Fig.3). We obtained anomaly high values on 190 days.

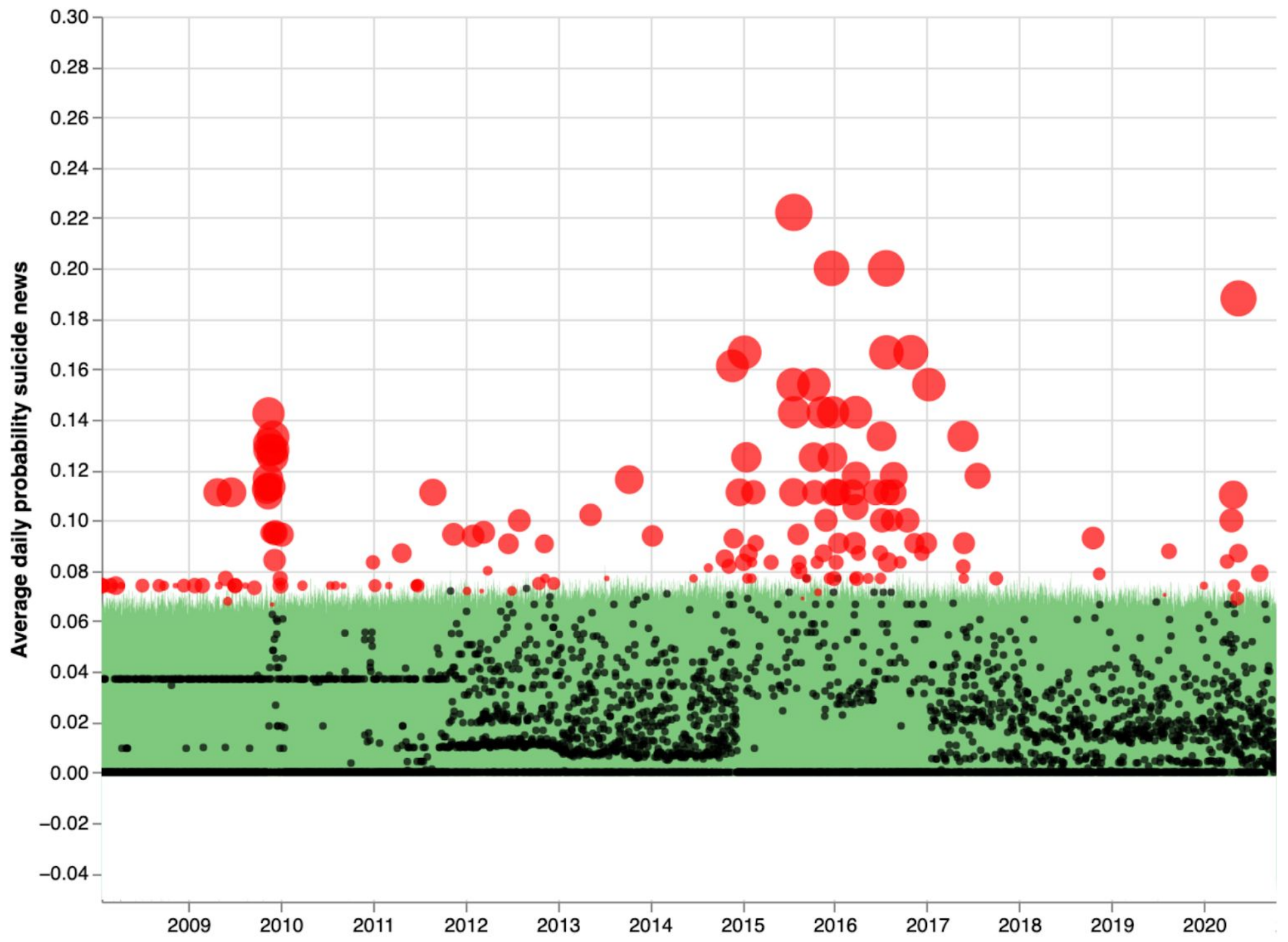

**Figure 3**. Anomalies detection. The temporal variation of the average daily probability of suicide contagion in news (black simbols) , the Porophet fitting curve (green lines), and the anomalies points (red symbols).

Table 1 shows the list of the most remarkable anomalies detected by the previous analysis (marked as red points in figure 3), the corresponding average suicide subject probabilities, and the event in the news. The match between the anomaly detection and a mediatic suicide cases confirms that the Prophet algorithm applied to our data is a good method to detect impacting suicide news.

| **Date** | **Suicide subject average probability** | **Event** |
|---|---|---|
| 2015-07-24 | 0.22 | The news refer to the death by suicide of a black woman in EEUU after being deteined.<br><br>https://elpais.com/internacional/2015/07/24/actualidad/1437730718_661809.html<br><br>https://www.abc.es/videos-otros/20150722/joven-negra-aparece-ahorcada-4369825582001.html |
| 2015-12-20 | 0.20 | The news cover a gang rape in which one of the five rapists died by suicide in jail.<br><br>https://www.publico.es/internacional/menor-involucrado-violacion-grupo-nueva.html<br><br>https://en.wikipedia.org/wiki/2012_Delhi_gang_rape_and_murder |
| 2016-07-23 | 0.20 | German-Iranian terrorist dies by suicide after killing nine<br><br>https://www.publico.es/internacional/atacante-munich-suicido-disparo-policia.html |
| 2016-11-29 | 1.66 | A father kills both his daughters and attempts suicide.<br><br>https://www.publico.es//sociedad//crimen-morana-abre-debate-cadena.html |
| 2017-05-24 | 0.16 | The police identifies the perpetrator of the suicide attack in manchester.<br><br>https://www.publico.es/internacional/ariana-grande-autor-atentado-manchester.html |
| 2009-12-03 | 0.13 | A mediatic case of a double morder with a scenified sucide is brought to justice.<br><br>https://www.abc.es/espana/comunidad-valenciana/abci-vladimir-ingresara-psiquiatrico-tras-condenado-anos-carcel-200912020300-1132288538565_noticia.html |
| | | https://www.abc.es/opinion/abci-vladimir-cumplira-pena-anos-psiquiatrico-200912020300-1132288565385_noticia.html |
| 2016-03-13 | 0.11 | The french Federal Bureau of Aircraft Accident Investigation concluded that the German Wings case was caused deliberately caused by the pilot, who died by suicide in the process taking the lives of 144 passengers and six crew members.<br><br>https://www.publico.es/internacional/medico-aconsejo-internar-lubitz-psiquiatrico.html |
| 2017-05-21 | 0.11 | A Spanish baker , a former chairman of the Spanish bank Caja Madrid, Miguel Blessa dies by suicide. In Feb. 2017 he was sentenced to a six-year jail term in connection with fraud events during tenure as Caja Madrid. He was connected to a corruption scandal involving political parties.<br><br>On 20 July 2017, he committed sucidided https://elpais.com/politica/2017/07/20/actualidad/1500535459_718664.html |

**Table 1.** List of recent anomalies, the corresponding average probabilities and the event

In order to visualize the daily anomalies of high suicide subject probabilities in more detail, we use a heatmap plot, using a warmer color for the high average suicide subject probabilities ( see Figure 4 which covers 2016 and 2018 years). Surprisingly, the high probabilities of suicide subject not only correspond to the anomaly day, but pop up over the course of several days, i.e. a relevant event remains on the news over time . In the studied case, this is reasonable, when an impacting suicide case occurs, this is talked about for several days (Figure 4). This is in accordance with our anomalie model, i.e. captures the likelihood of suicide news.

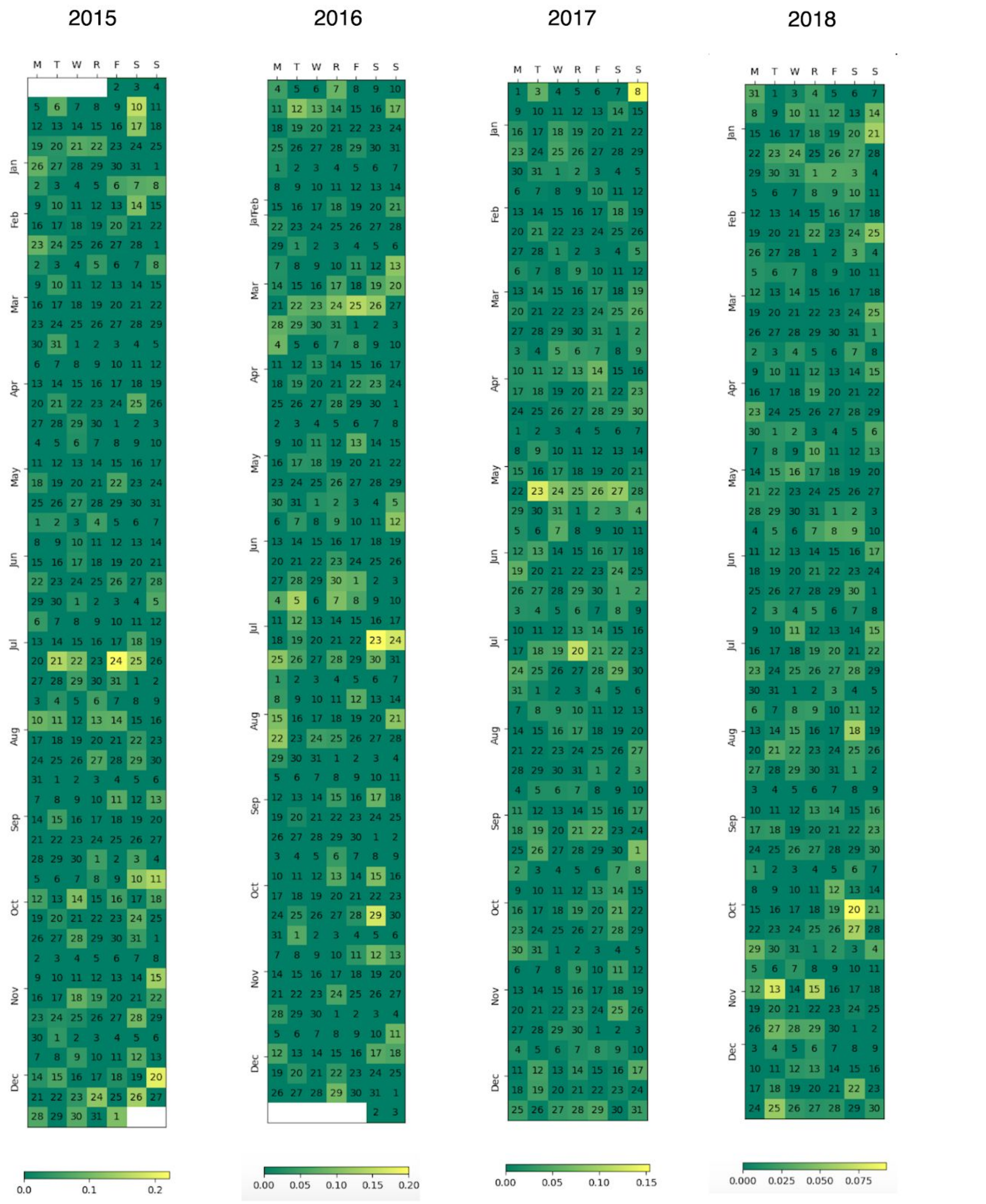

**Figure 4**. Suicide subject news heatmap from 2015 to 2018. Each day is depicted with a color according to the probability of suicide news (see legend).

## 4. DISCUSSION

### 4.1 Intrinsic thematic relationship of suicide news

In the following, we focus on the intrinsic connection of subjects (or tags) in the news. We use the extracted suicide subject probability (outputted by the neural network NN2) to discern which news talks about this issue. We extract how likely is that each news address suicide topics by a suicide subject binary classification (in NN2, see Fig. 1b). We develop a neural network composed of an embedding layer, four convolutional (1 dimensional) layers, two poolings and one output dense layer (with dimension equal to the number of possible tags). In this case, we obtain binary cross entropy loss 0.0233 (accuracy of 99.91%, precision of 99.91 and recall of 99.85%). This model is trained using the suicide tweets and non suicide tweets as input data training. The model associates each text input with a probability of suicide subject, that is, a 0-1 range where zero means lackness of suicide subject.

We consider a new topic is realted with suicide subject if the probability returned by the neural network is greater than 0.999, i.e., the neural network NN2 confirms it without doubt. This process leaves us with a set of 9160 news that cover suicide in this way. Figure 5 shows these related topics group by year. The most common tags related with suicidade are islamism (29.21%), culture (18.91%) and politics (14.53%) (Fig.5 right). The tag culture and politics correspond respectively to celebrities and politicians, which committed suicidade. While islamism mainly correspond with suicide terrorims as a tactic.

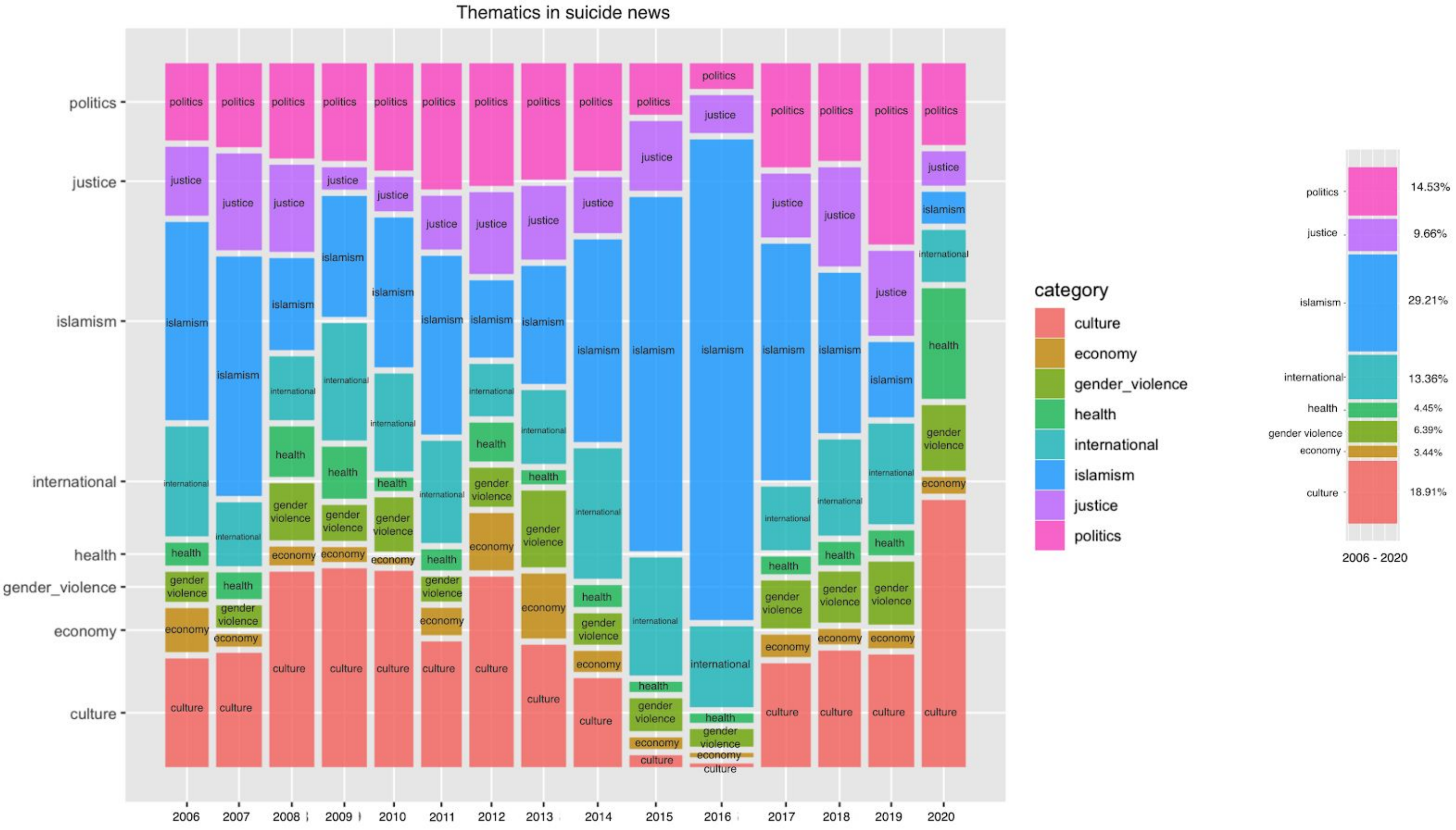

**Figure 5**. Categories (tags or wide thematics) of suicie new group by year (left) and sum(right). Note that the year 2020 only includes data from January to October.

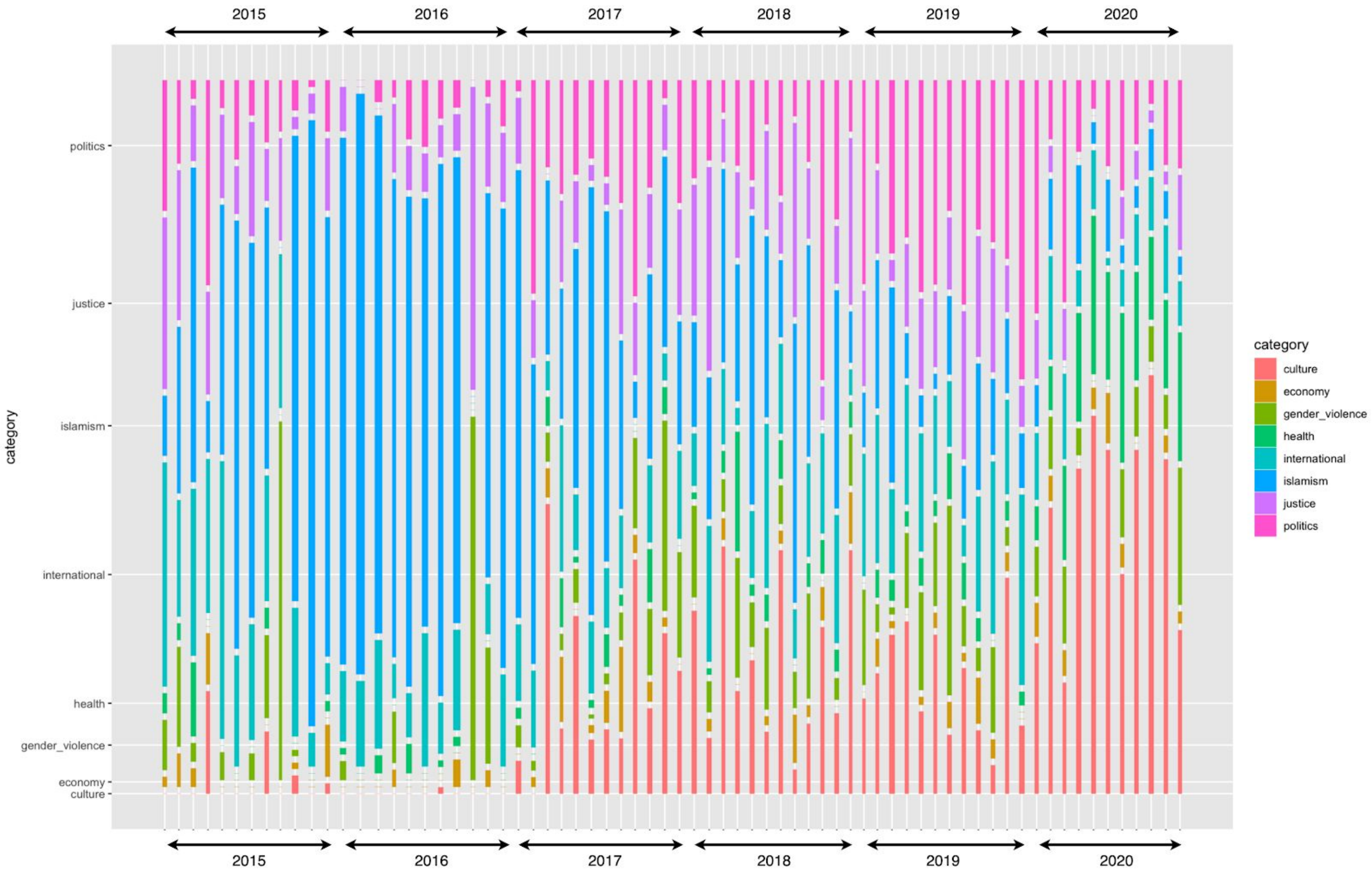

**Figure 6**. Categories (tags or wide thematics) of suicie new organized by months from 2015 to 2020

Despite that over the years the main topics which a suicidade connotation do not change, the percentage of them does. Exemplarily, islamism tag is most relevant in 2006 -Charlie Hebdo published Mahoma Comic-, 2012-2016 -islamic suicidade attacks in Europe-. In 2020, the tag *health* enhanced its relevances, it may be related with the Coronavirus Pandemic (Gunnell et al., 2020; Halford et al., 2020). As shown in Fig. 6, where we repeat the analysis by monthly averages, there is a significant increment of the health topic in April 2020. In the first semester 2017, we see an increase of suicide news in the topic is *politics,* which corresponds with the event shown in the anomalies analysis (see Table 1) the judicial process of the corruption politician scandal and the suicidade of Miguel Blesa. Remarkably, within 2018-2019, gender violence topic increases, which may be explained with the awareness of Gender Based violence and the boost of feminicides that end with the wuicide of the predator.

One key part of our analysis in the Topological data analysis (TDA) algorithm called Mapper. Since TDA is a relatively new branch of data analysis, we will introduce some of its principles and ideas. Topology is regarded as a pure or theoric branch of mathematics. The fundamental assumption in topology is that the way in which the parts of a system are connected is more important than distance. This particular approach is important to why topology is useful for analyzing and visualizing data. Topological Data Analysis is a modern discipline that uses the techniques of topology to analyze data. Mapper is a TDA algorithm that gives us a way to construct a graph (or simplicial complex) from data in a way that reveals some of the topological features of the space. It was developed by (Singh et al., 2007). Though not exact, mapper is able to estimate important connectivity aspects of the underlying space of the data so that it can be explored in a visual format. We used the python implementation of the Mapper algorithm Kepler Mapper to study the connectivity of suicide news and the rest of the news within a period (Veen et al., 2019).

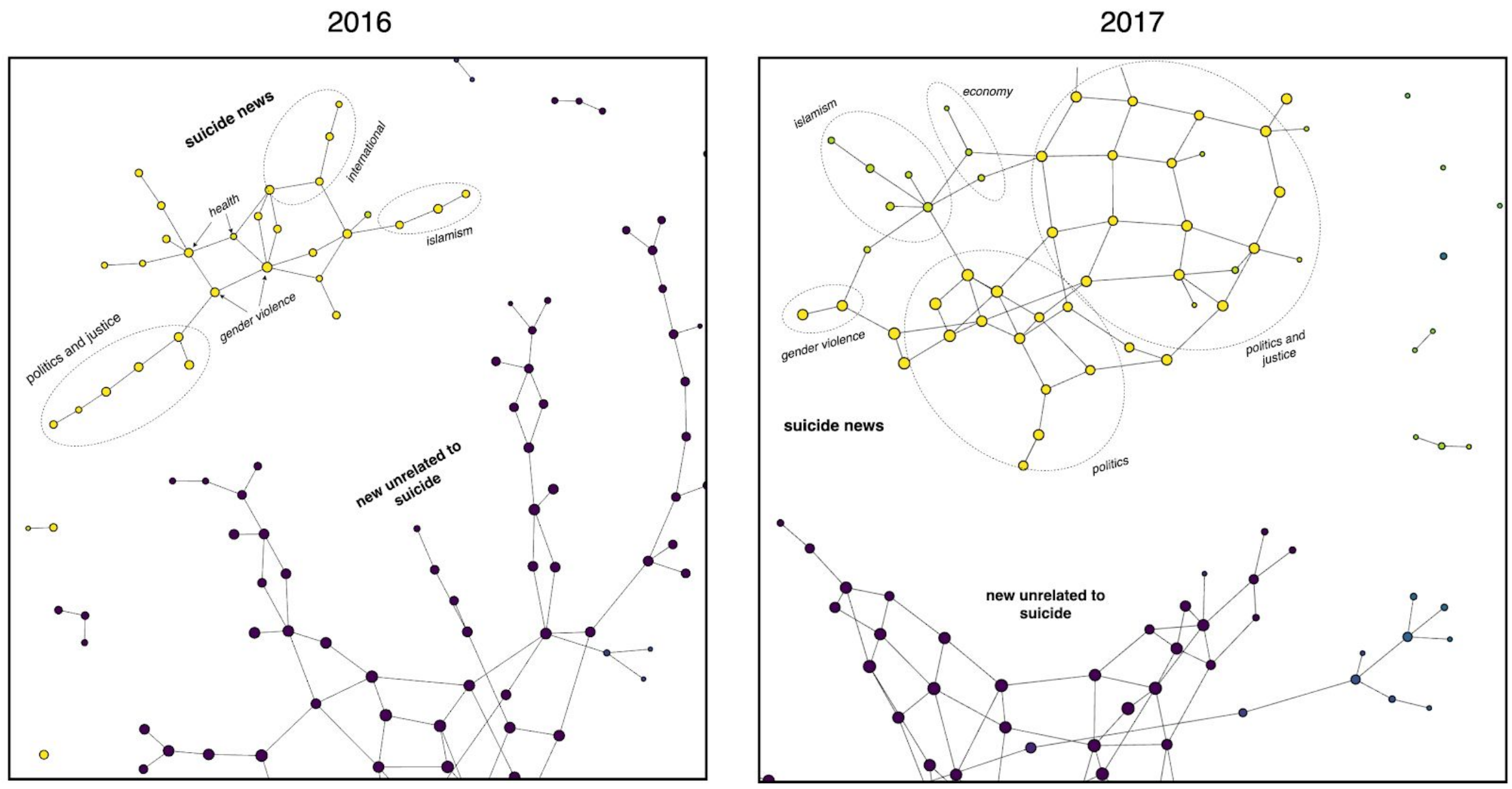

**Figure 7**. **Interdependence between topics.** Subject dependence. (left) 2018 and (right) 2019 diagram. Each node represents a cluster of news. Note that news belongs to the same cluster if they are closely related in terms of their subjectsI.e. the algorithm then connects clusters if they overlap. The colour of the nodes corresponds to the suicide subject probability (the higher suicidade probability, the warmer colour is). The size of the nodes represents the amount of news in them.

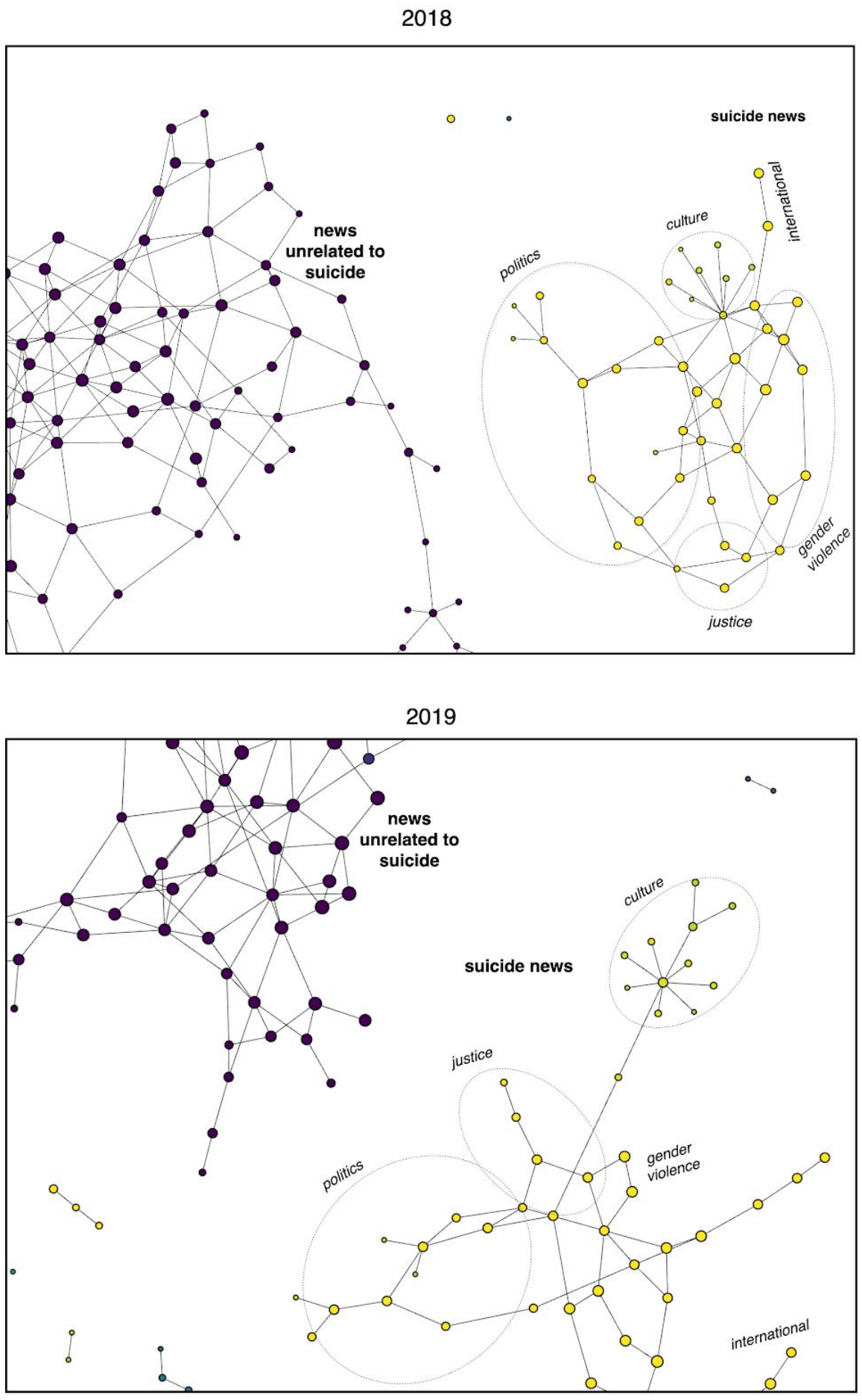

**Figure 8. Interdependence between topics.**.Constructed using the same setting as Figure 7. 2018 up and 2019 right.

The diagram returned by Mapper algorithm is obtained following this process. First, we restrict the dataset selecting the news for a given year. For each news, we obtain the sequence of probabilities of each tag (using the subject classification neural network NN1), which gives us a sequence of vectors of more than 1000 dimensions (one for each possible tag). Second, we reduce the dimensionality of the previous sequence of vectors using principal component analysis. Then, we obtain a sequence of vectors (one for each new) with dimension 3 (the reduced dimension). This new sequence of vectors captures the geometric information of the thematic similarity of the news. Thirld, we apply the Mapper algorithm (Veen et al., 2019), which summarizes the closeness of the news with a geometric interpretation. The resulting graph draws a node for each cluster of news where the ones that are thematically close in terms of their subjects. Note that if these nodes are overlapped, i.e. there are news belonging to more than one cluster, the nodes will be joined with segments. This will allow us to see how subjects change and how news organize themselves in terms of closeness. Figure 6-7 show the corresponding plot, where the warmer colors represent news (or clusters of them) with higher suicide subject probability. we check the branches of the graph. Each branch represents news' clusters whose subjects evolve similarly. In the 2016 diagram (figure 7, left), we see how there is a big branch of suicide news which are related with politics and justice that later merges with domestic violence news and health news. Such analysis evidences the entanglement between all these topics which, at a first glance, seem uncorrelated.

Indeed, there is a connection between international and domestic violence which could be related with the #metoo movement (Kero et al., 2020). Hence, all these connections appear in all the studied years (see figures 7 and 8) .

4.2 Conclusions

A strength of the developed is Machine Learning study is the suicidade topic extraction of the media articles and the definition of a *suicide subject probability for each new*. This suicide subject probability technique is especially fruitful when we use it to map all news within a period, which pinpoints several events that pushed both the increase in news regarding this subject and public awareness in Spain (table 1 and figure 4). Our research direction is nevertheless one of many, here we have focused on extrapolate the effect of mediatic suicide cases in the reporting of suicide news.

Based on the hypothetical protective effect and in order to prevent contagion (Papageno effect), the WHO (World Health Organization) developed guidelines and recommendations for the news media professionals to ameliorate the elaboration of news on suicide (*WHO | Preventing Suicide*, n.d.-b; *WHO*, n.d.) supported by other associations dedicated to suicide prevention (*IASP - Special Interest Group: Suicide and the Media - Guidlines*, n.d.; *WHO*, n.d.). This strategy is applied to the general population, where few strategies have been proved effective (Turecki & Brent, 2016; Zalsman et al., 2016). On general lines, journalists should avoid explaining and detailing suicidal methods and should focus on the possibility to get help and overcome the suicidal crisis, with the aim to educate the population in mental health issues (*WHO | Preventing Suicide*, n.d.-a).

Finaly, our approach shows that suicide tends to be connected by the media with the subject "*politics", "justice", "culture", "Islamism" and "gender violence"* and rarely with the subject "*health"* (figures 5,6 and 7). This finding may set a new point of view and representation of the news about suicide that would lead to better reporting. On this line, creation or improvement of already existing guidelines could be an interesting objective.  Hence, news media would help raise awareness on the problem that suicide poses, and thus, promote resources for its prevention.

**Data availability**

The data that support the findings of this study are available from the corresponding author upon reasonable request.

**Code availability**

The source code used in this study is available at
https://github.com/news-scrapers/news-scraper-subject-classifiers-model
and
https://github.com/news-scrapers/suicide-subject-classifier-neural-networks-models

**Corresponding author**

Correspondence to Hugo J. Bello.

**Competing interests**

The authors declare no competing interests.